# AMDNet23: A combined deep Contour-based Convolutional Neural Network and Long Short Term Memory system to diagnose Age-related Macular Degeneration

Md. Aiyub Ali[1], Md. Shakhawat Hossain[1], Md.Kawar Hossain[1], Subhadra Soumi Sikder[1], Sharun Akter Khushbu[1] and Mirajul Islam[1]

[1] Department of Computer Science and Engineering, Daffodil International University, Dhaka 1341, Bangladesh

Correspondence: Mirajul Islam; merajul15-9627@diu.edu.bd

**Abstract**

In light of the expanding population, an automated framework of disease detection can assist doctors in the diagnosis of ocular diseases, yields accurate, stable, rapid outcomes, and improves the success rate of early detection. The work initially intended the enhancing the quality of fundus images by employing an adaptive contrast enhancement algorithm (CLAHE) and Gamma correction. In the preprocessing techniques, CLAHE elevates the local contrast of the fundus image and gamma correction increases the intensity of relevant features. This study operates on a AMDNet23 system of deep learning that combined the neural networks made up of convolutions (CNN) and short-term and long-term memory (LSTM) to automatically detect aged macular degeneration (AMD) disease from fundus ophthalmology. In this mechanism, CNN is utilized for extracting features and LSTM is utilized to detect the extracted features. The dataset of this research is collected from multiple sources and afterward applied quality assessment techniques, 2000 experimental fundus images encompass four distinct classes equitably. The proposed hybrid deep AMDNet23 model demonstrates to detection of AMD ocular disease and the experimental result achieved an accuracy 96.50%, specificity 99.32%, sensitivity 96.5%, and F1-score 96.49.0%. The system achieves state-of-the-art findings on fundus imagery datasets to diagnose AMD ocular disease and findings effectively potential of our method.

**Keywords:** AMDNet23, Fundus image classification, CNN-LSTM, ocular diseases, automated diagnosis, convolutional neural networks, long short-term memory, early detection, Medical imaging, diagnosis.

**Introduction**

Over the past two decades, ocular diseases (ODs) that can cause blindness have become extremely widespread. ODs encompass a wide range of conditions that can affect various components of the ocular, including the corneal tissue, lens, retina, optic nerves and periorbital tissues. Ocular diseases include abnormalities such as cataracts, untreated nearsightedness, trachoma, macular degeneration associated with aging, and diabetes-associated retinopathy. These ailments play a substantial role in global retinal degeneration and visual impairment. [1]. The worldwide prevalence of near- or farsightedness vision deficiency affects over 2.2 billion individuals [2]. Approximately half of the total cases, amounting to at least 1 billion folks, as reported by the World Health Organization (WHO), suffer from vision impairments that could have been evaded or remain unattended. Among these individuals, around 88.4 million have untreated refractive errors leading to adequate to extensive distant impaired vision, nearly ninety-four million owned cataracts, and eight million individuals are possessed by aged-related macular degeneration, diabetic retinopathy (3.9 million). [3] Despite significant investment, the number of individuals living with vision loss might increase to 1.7 billion by 2050, from the 1.1 billion people accomplished in the year 2020. Age-associated macular degeneration (AMD) predominantly strikes the older demographic, resulting in the gradual deterioration of the macula, a crucial part of the retina in charge of the central region of perception. The consequences of AMD manifest as central vision abnormalities, including blurred or distorted vision, which significantly impede various daily activities[4].

Accurate and earlier identification of AMD disease explicitly a vital role in safeguarding irreversible damage to vision and initiating timely treatment and safeguarding ocular health. Machine learning techniques have advanced to the point where early identification of aged macular degeneration eye illness by an automated system has significant advantages over manual detection [5]. As aids in diagnosing eye diseases, digital pictures of the eye and computational intelligence (CI)-based technologies assist doctors in diagnosis. In the realm of diagnosing eye diseases, digital eye images and computational intelligence (CI)-based technologies serve as indispensable tools, enabling doctors to enhance their diagnostic capabilities[6]. In medical imaging, there are also various approaches are employed including fundus photography, [7] optical coherence tomography (OCT), and imaging modalities specifically designed for the eye. These imaging technologies allow for detailed visualization and analysis of ocular structures, facilitating the identification of characteristic features and abnormalities associated with age-related macular degeneration diseases.

Several researchers have demonstrated a critical task in ophthalmology, facilitating the early detection and diagnosis of aged macular degeneration (AMD) ocular disease using fundus image.

Researchers have focused on deep learning [8-10], The fields of vision for computing [11,12] and the use of predictive learning of machines [13,14] method have used to develop robust classification models to identify retinal images into AMD disease categories accurately. The incorporation of deep learning methodologies [57,58] plays a pivotal role in accurately classifying diverse ocular diseases, thereby ensuring the advancement of intelligent healthcare practices in the field of ophthalmology [15].

Therefore, this paper seeks to demonstrate a novel system employing a AMDNet23 framework, the deep mechanism of CNN and LSTM networks is combined for the automated identification of AMD from fundus photography. Within this approach, CNN performs the purpose of extracting fundus features, and LSTM undertakes the crucial task of classification AMD constructed on the extracted features. Internal memory inside the LSTM network empowers it to learning knowledge gained from significant experiences with extended period of condition. In the fully interconnected networks, each layer is linked comprehensively, and the nodes in between layers construction are unconnected, and LSTM nodes connection within a directed graph accompined a temporal order, which serves as an input with a specific form [16] . The hybrid two dimensional CNN and LSTM system combination improves classification of AMD ophthalmology and assists clinical decision, the dataset collected from several sources and preprocessing technique for the image quality enhancemnet to classify AMD efficiently. The assets of this research have been articulated in the following.

a) Constructing a combination of CNN-LSTM based AMDNet23 framework for the automated diagnosis of AMD and aiding clinical physician in the early detection of patients.
b) The collection data has investigated by employing the contour-based quality assessment technique in identifying the structure of fundus photography, Ocular illumination levels fundus images are automatically eliminated.
c) To enhance image quality, CLAHE improves the visibility of subtle details and enhances local contrast and Gamma correction adjusts the intensity levels, improving image quality and facilitating better diagnosis of AMD.

d) AMDNet23 hybrid framework for detection of AMD utilizing fundus image ophthalmology, data comprising 2000 images equitively.
e) An empirical evaluation is accessible encompassing accuracy, specificity, sensitivity, F1-measure, and a confusion matrix to assess the effectiveness of the proposed method.

The rest of the contents of this article are arranged a manner as follows: Section II covers the related works of this research. Section III Section III articulates the proposed AMDNet23 methodology, including data collection and preprocessing techniques, and a comparison of some existing models. Section IV covers the experimental findings and discussion, including state-of-the-art and transfer learning comparisons. The conclusion is presented in Section V.

**Related work**

In the pursuit of identifying the ocular disease, researchers have harnessed the power of deep learning techniques. These methods leverage fundus ophthalmology to facilitate the diagnosis of ocular diseases. This reviewed literature presents cutting-edge systems that developed deep-learning techniques for detecting AMD, diabetes, and cataracts.

M Sahoo et. al[17] proposed an innovative ensemble-based prediction model called weighted majority voting (WMV) for the exclusive diagnosis of Dry-AMD. This approach intelligently combines the predictions from various base classifiers, utilizing assigned weights to each classifier. The WMV model demonstrates remarkable accuracy, achieving 96.15% and 96.94% accuracy rates, respectively. P Muthukannan et. al. [18] introduced a computer aided approach that leverages the Flower pollination optimization approach (FPOA) in accompanied with a CNN mechanism for preprocessing, specifically utilizing the maximum entropy transformation on the ODIR public dataset. The model's performance was then benchmarked against other optimized models, demonstrating superior accuracy at 95.27%. In a study by Serener et al. [19], their goal was to employ OCT images and deep neural networks to detect both dry and wet AMD. Regarding the purpose, two architectures—AlexNet and ResNet—were used. The outcomes revealed that the eighteen-layer ResNet model correctly identified AMD with an astounding accuracy of 94%, whereas the AlexNet model produced an accuracy of 63%.

There are several deep learning methods for detecting cataracts, because of the drawbacks of feature extraction and preprocessing, these methods don't always produce adequate results. Kumar et al. [20] proposed several models to improve clinical decision-making for ophthalmologists. Paradisa et al. [21] Fundus images applied the Concatenate model with For feature extraction, Inception-ResNet V2 and DenseNet121 are implemented, and MLP is deployed for classification and average accuracy was 91%. Faizal et al. [22] An automated cataract detection algorithm using CNN achieves high accuracy (95%) by analyzing visible wavelength and anterior segment images, enabling cost-effective early detection of various cataract types. Pahuja et al. [23] To enhance the model performance, data augmentation and methods to extract features have been performed. Therefore they used CNN and SVM models for the detection of cataract on a dataset comprising normal and cataract retinal images, achieving high accuracy of 87.5% for SVM and 85.42% for CNN. Hence, et al. [24] used a CNN model to diagnose cataract pathology with digital camera images. The model achieves high accuracy (testing: 0.9925, training: 0.9980) while optimizing processing time. It demonstrates the potential of CNNs for cataract diagnosis. Although et al [25] used color fundus images to detect cataracts.

A variety of computer vision engineering approaches are used to forecast the Diabetic retinopathy (DR)'s occurrences and phases automatically. Mondal et al. [26] Their model is a collaborative deep neural system for automated diabetes-related retinopathy (DR) diagnosis and categorization using two models: modified DenseNet101 and ResNeXt. Experiments were conducted on APTOS19 and DIARETDB1 datasets, with data augmentation using GAN-based techniques. Results show higher accuracy with accuracy for each of the five classes reached 86.08%, whilst for each of the two classes the score was 96.98%. Whereas, a ML-FEC model with pre-trained CNN architecture was proposed for Diabetic Retinopathy (DR) detection using ResNet50, ResNet152, and SqueezeNet1. On testing with DR datasets, ResNet50 achieved 93.67%, SqueezeNet1 achieved 91.94%, and ResNet152 achieved 94.40% accuracy, demonstrating its suitability for clinical implementation and large-scale screening programs. Using a novel CNN model, Babenko et al. [27] were able to multi-class categorize retinal fundus pictures from a publically accessible dataset with an accuracy of 81.33% for diabetic eye disease. Priorly based on UNet architecture, et al. [28] achieved 95.65% accuracy in identifying red lesions and 94% accuracy in classifying DR levels of severity. The approach was examined utilizing publically accessible datasets: IDRiD (99% specificity, 89% sensitivity) and MESSIDOR (94% accuracy, 93.8% specificity, 92.3% sensitivity).

## Methodology

Aging macular degenerative disorder (AMD) is an advancing retinal condition that predominantly impacts individuals over the age of 50. This eye disease can significantly affect eyesight, leading to various visual issues like blurred or distorted vision, where straight lines might appear wavy or twisted. Moreover, it causes a loss of central vision, the emergence of dark or empty spots at the center of vision, and alterations in color perception. Thus taking proactive steps to prevent eye diseases is crucial for maintaining clear and vibrant sight throughout our lives. In recent years, neural networks containing layers of convolution (CNNs) have demonstrated considerable potential in the processing of medical images. It has the remarkable capacity to recognize and extract meaningful features from images automatically. Figure 1 outlines the steps in developing the proposed CNN-based methods for AMD eye disease detection:

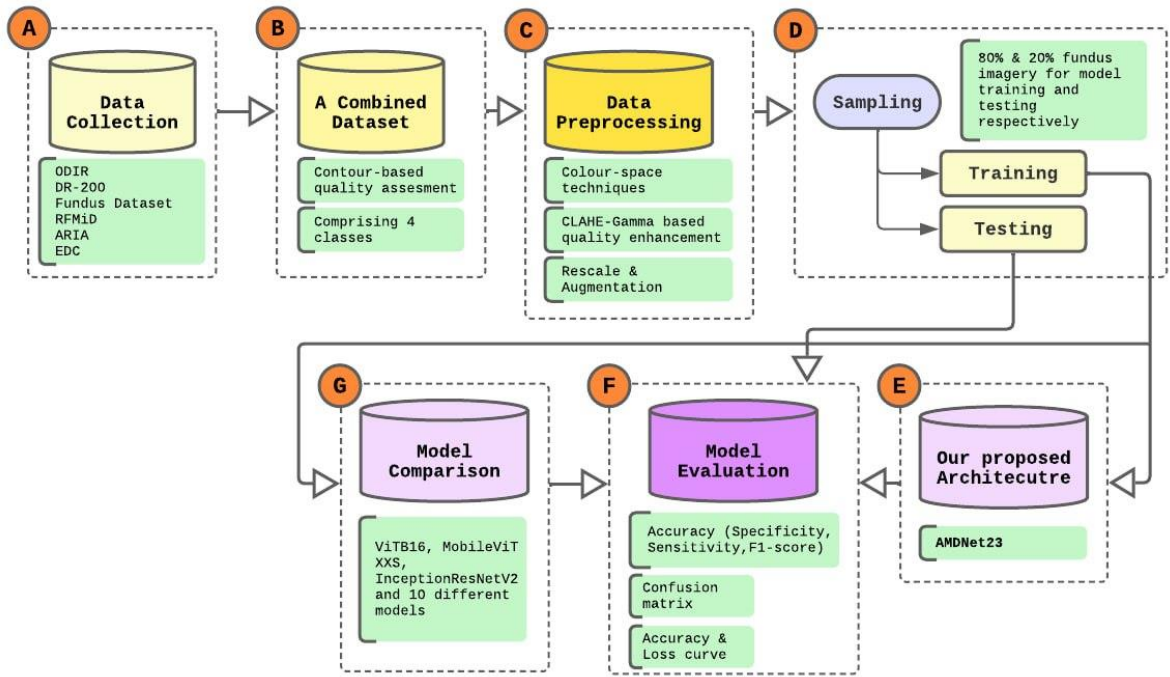

Figure 1: Overall proposed-based Method

**A. Data Collection:**

The case of normal class represents the absence of any specific eye disease or condition. A healthy eye functions optimally, providing clear and unimpaired vision. Diabetes, a systemic disease characterized by elevated blood sugar levels, can lead to various ocular complications. Diabetic ocular disorders, notably diabetic retinopathy, occur when the blood vessels found in the retina undergo damage as a consequence of elevated blood sugar concentration [29]. Older individuals are predominantly affected by age-associated macular degeneration (AMD) and involves the progressive deterioration of the macula, a small but crucial of the core vision-related region of the retina. AMD can lead to blurred or distorted central vision, impacting activities [30]. A cataract is another common eye condition, particularly associated with aging. It involves clouding the crystalline lens inside the eye, leading to inconsistent or foggy vision [31]. Fig. 2. Shows the sample images of Normal, Cataract, AMD and Diabetes respectively.

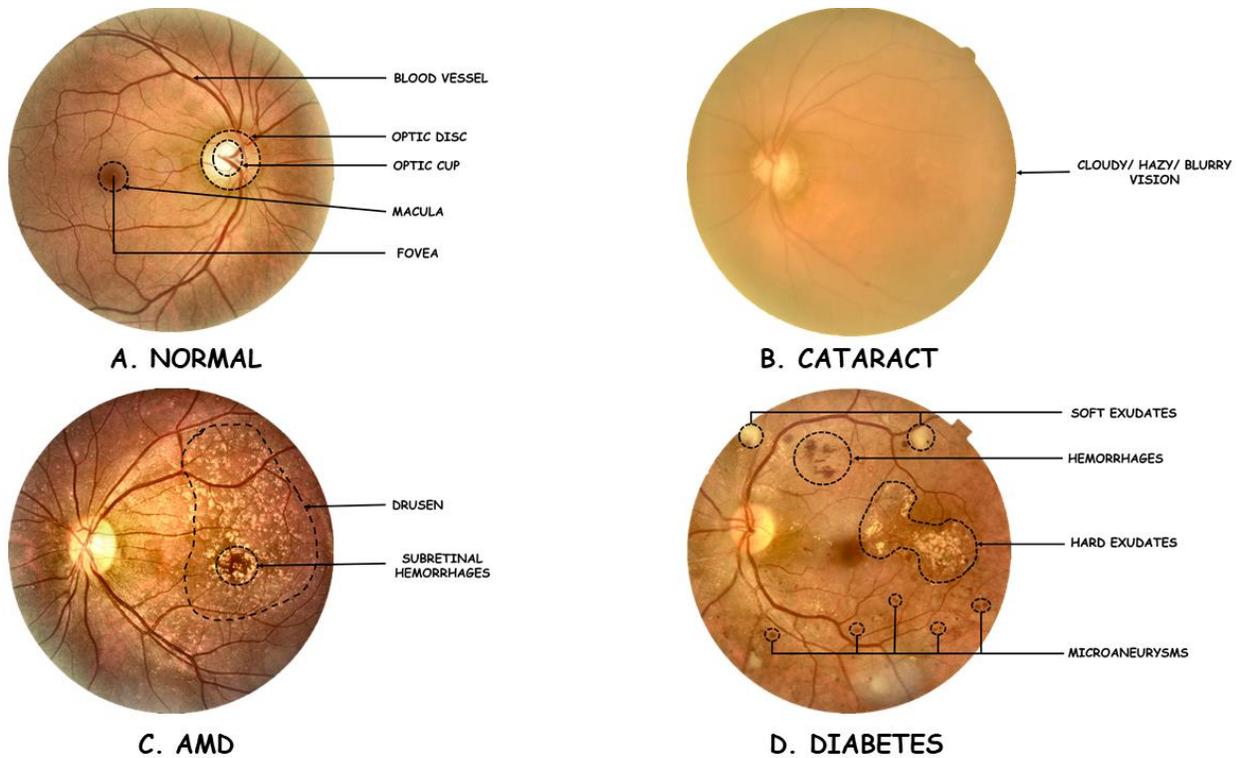

Figure 2: Types of fundus ophthalmology

To train a robust CNN model, a diverse and well-annotated dataset of AMD and non-AMD eye images is essential. The dataset employed in this study containing a total of 2000 images, was put together by assessing the quality of the images from six other public datasets. Those datasets are ODIR[32], DR-200[33], Fundus Dataset[34], RFMiD[35], ARIA[36], and Eye_Diseases_Classification[37]. From these datasets. The quality assessment was done using contour techniques[38]. The contour-based approach focuses on the sharpness and clarity of edges, as they play a crucial role in human assess the quality of imagery. Which included illumination level, visibility structure, color and contrast, and direct eye image. Figure 3 represents a sample of the assessed image quality.

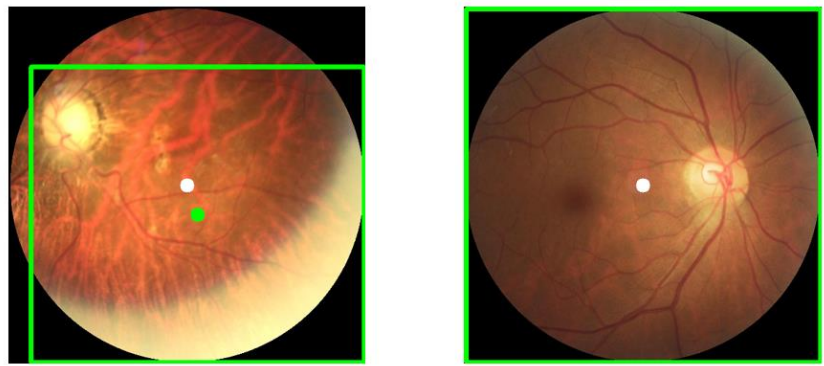

Figure 3: Contour-based approach

Here it can be seen that sharp, well-defined edges contribute to high-quality images, while blurry or distorted edges indicate poor quality. Such poor-quality images would negatively impact the machine's perception. We put together a dataset consisting of four classes: Normal, Diabetes, AMD, and Cataract, where each class contains 500 images. Table 1 indicates the quantity of accessible and selected images (within the first bracket) from those six public datasets.

| Datasets | Normal | Diabetes | AMD | Cataract |
|---|---|---|---|---|
| ODIR[33] | 2873(168) | 1608(300) | 266(266) | 293(256) |
| DR-200[34] | 1000(332) | 1000(150) | | |
| Fundus Dataset[35] | | | 46(46) | 100(44) |
| RFMiD[36] | | 376(50) | 100(100) | |
| ARIA[37] | | | 101(88) | |
| Eye_Diseases_Classification[38] | | | | 1038(200) |

**B. Data Pre-processing:**

Preprocessing, which is the strong suits of the proposed work, was focused on enhancing image quality, and some of the preprocessing techniques applied to this work were not used by the previously proposed cataract disease detection works. The data were preprocessed in different color spaces (as shown in Figure 4) for extracting the features while bettering the practicability of our models. Among the RGB(G), HSV(V), and LAB(L) color spaces, the vessels were visible in the LAB(L) color space. As a result, the LAB(L) color space was chosen.

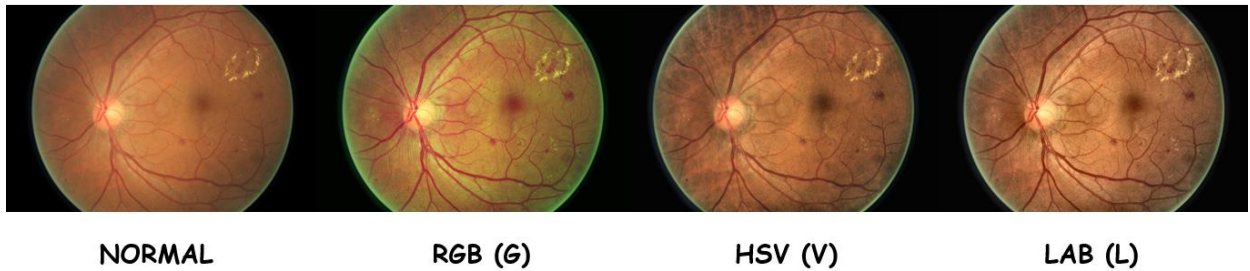

Figure 4: Color spaces

Later, some preprocessing algorithms like CLAHE and Gamma correction[39] were applied to enhance image quality by adjusting the brightness and contrast. We experimented with several parameters for these algorithms and finally got the satisfying result for CLAHE(2.0, (8,8)). For gamma values of 0.5, the image is found to become darkened. Moreover, for gamma values of 2.0, the image is found quite faded. To overcome this problem CLAHE is used to enhance regional contrasting, making the image more visually appealing and informative[40]. Figure 5 shows the resulting images for our experimented algorithms along with the finalized CLAHE (2.0, (8,8)) for our model.

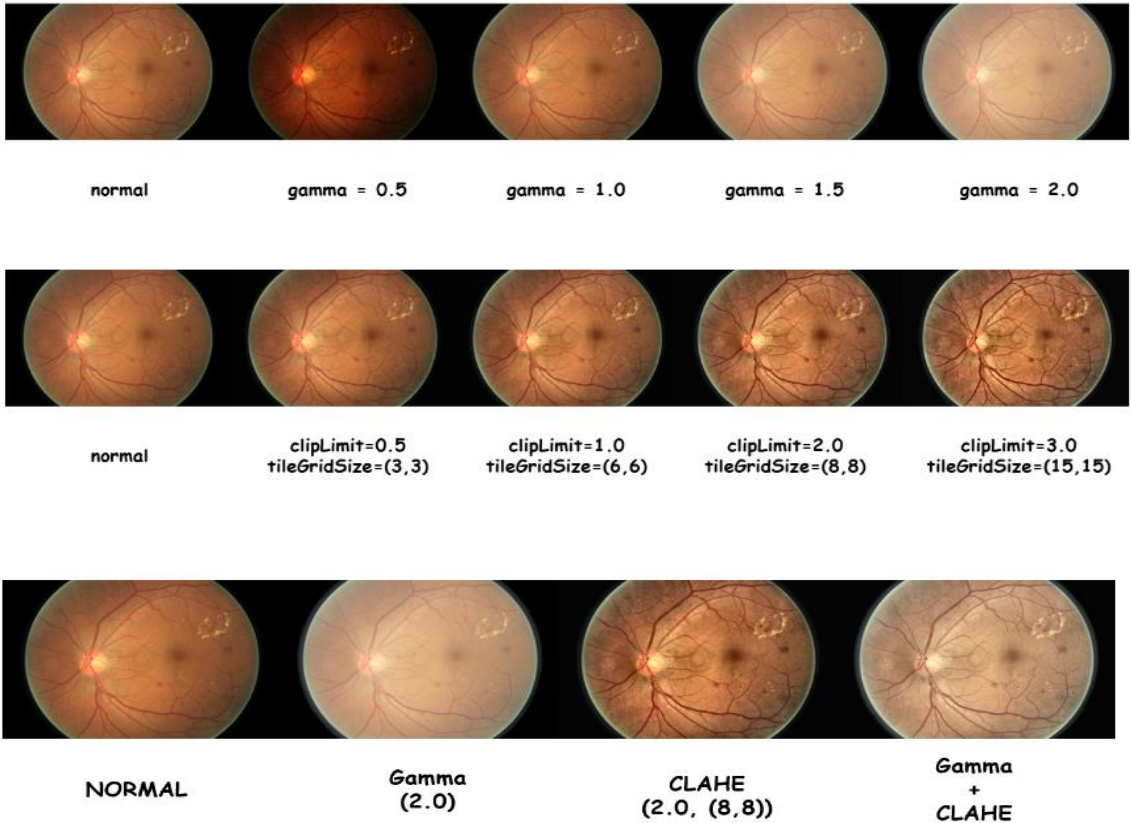

Figure 5: Gamma and CLAHE based quality enhancement

Histogram comparison between the applied Context-limited adapted equalization of histograms (CLAHE) preprocessing algorithm and a normal image in Figure 6 can help illustrate the effects of CLAHE on enhancing local contrast and improving image quality.

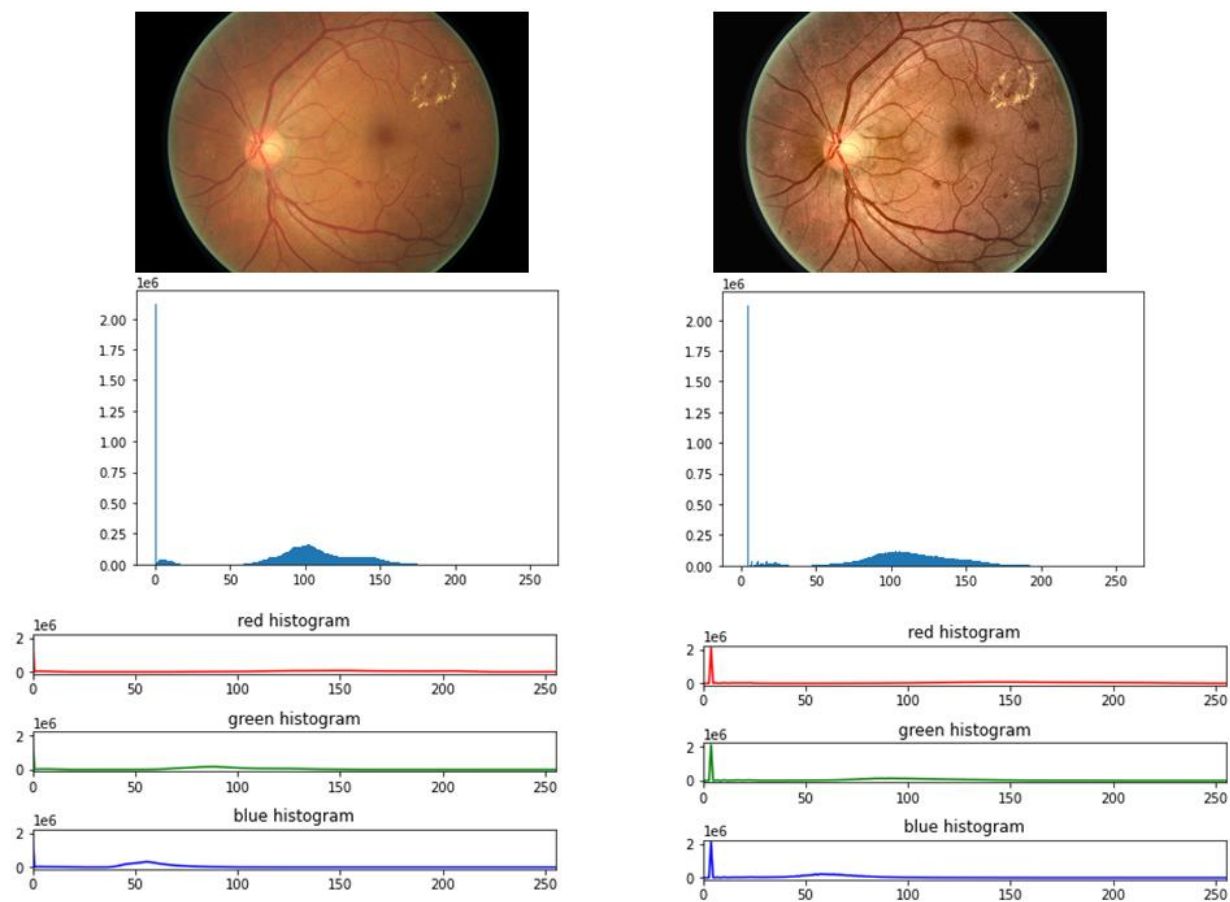

Figure 6: Hitogram comparisons

By comparing the histograms, we can observe the changes in pixel intensity distribution before and after applying CLAHE. In the normal image, the histogram exhibits a relatively uniform distribution of pixel intensities, with some variations depending on the content of the image. CLAHE is implemented as a component in the preprocessing, it adapts the contrast enhancement locally, making it particularly effective in improving the contrast of regions with varying intensities. This helps reveal hidden details and textures that might have been obscured in the original image.

| **Images** | **MSE** | **PSNR** | **SSIM** |
| --- | --- | --- | --- |
| 2376_left.jpg | 2388.13 | 14.35 | 0.48 |
| 84_right.jpg | 2189.98 | 14.72 | 0.65 |
| 980_right.jpg | 927.54 | 18.45 | 0.53 |
| 71_left.jpg | 709.37 | 19.62 | 0.54 |

The effectiveness of the quality of image preprocessing, Table 2 displays the readings of the metrics mean square errors (MSE), Peak Signal-to-Noise Ratio (PSNR), and Structures Similarity Index (SSIM) [41].

These metrics compare the preprocessed image to the original image to determine the level of distortion or similarity.

**Mean Squared Error (MSE):** MSE generates the resultant mean squared disparity between the preprocessed and original image's pixel values. Lesser MSE reading indicate greater similarity between the images. MSE is calculated using the formula:

$$MSE = \frac{1}{m*n} \sum_{x=0}^{m-1} \sum_{y=0}^{n-1} [I(x,y) - K(x,y)]^2$$

where $m*n$ represents the image dimensions, $I(x,y)$ and $K(x,y)$ indicate preprocessed image's pixel values and original images at coordinates $(x,y)$.

**Peak Signal-to-Noise Ratio (PSNR):** The optical appealing of preprocessed photographs is frequently assessed utilizing the PSNR measure. It calculates a measure of the peak power ratio of the signal's strength to noise, which assessed in decibels (dB). Increased PSNR values indicated a greater similarity between the images. PSNR is calculated using the formula:

$$PSNR = 10 log_{10} \frac{MAX^2}{MSE}$$

wherein MAX is the highest pixel value that is permitted to be used (for example, 255 in 8-bit photographs).

**Structural Similarity Index (SSIM):** SSIM evaluates the luminosity, contrary, and structural similarities between the preprocessed image and original image. The value of 1 denotes complete similarity, using SSIM readings varying from -1 to 1. Higher SSIM values indicate better similarity between the images. SSIM is calculated using a combination of mean, variance, and covariance of the image patches.

$$SSIM = \frac{(2\mu_x \mu_y + c1)(2\sigma_{xy} + c2)}{(\mu_x^2 + \mu_y^2 + c1)(\sigma_x^2 + \sigma_y^2 + c2)}$$

where c1 and c2 are constants to prohibit division by zero, σ and μ stand representing the standard deviations and mean respectively.

By calculating MSE, PSNR, and SSIM before and after image preprocessing, The effectiveness of the preprocessing techniques in preserving image quality or reducing noise, artifacts, or other undesired effects can be determined. Lesser MSE readings, greater PSNR readings, and greater SSIM values indicate better image quality preservation.

### C. Comparison of some existing models

**Transfer Learning:**

In this research study, a handful of models were trained and evaluated. Some of those models are discussed below in the following sections:

**i.ViTB16**

The ViTB16 model, also known as Vision Transformer Base with a depth of 16 layers, is a deep learning architecture specifically designed for image classification tasks [42]. 224*224 pixels were made up of the size of the input images. A grid of patches containing fixed sizes is used to divide the input image. Each individual patch is positioned linearly to obtain a lower-dimensional representation. The patch embeddings are enhanced by employing positional encoding to provide the model with spatial information. The model is capable of finding relationships dependencies between different patches thanks to a self-attention mechanism. It calculates attention scores between all pairs of patches and applies weighted averaging to aggregate information. Layer normalization is applied after the self-attention mechanism to normalize the output and improve training stability. SGD (Stochastic Gradient Descent) optimizer was employed for training the model, and the learning rate was 0.0001.

**ii. DenseNet121, DenseNet169:**

DenseNet[43] is a famous deep-learning architecture known for its dense connections between layers, enabling effective feature reuse and alleviating the vanishing gradient problem. DenseNet121 and DenseNet169 have 121 and 169 layers, respectively, making DenseNet121 a relatively shallow variant than DenseNet169. DenseNet121 has fewer parameters compared to DenseNet169, which makes it more memory-efficient and faster to train. DenseNet121 performs well on various image classification tasks but may not capture as fine-grained features as deeper models. DenseNet169 performs better than DenseNet121, especially when the dataset is larger and more complex. Choosing between DenseNet121 and DenseNet169 for a particular purpose like AMD classification, it is essential to consider the size and complexity of the dataset. Since the dataset used in this study was small, DenseNet121 should have been the model to pick, but we experimented with all DenseNet variants.

**iii. InceptionResnetV2**

A powerful convolutional neural network conception that incorporates the Inception and ResNet modules is termed the InceptionResNetV2 system [44]. It was proposed as an extension to the original Inception and ResNet models designed to improve image classification efficient tasks. The InceptionResNetV2 model was initialized with pre-trained ImageNet weights, excluding the top classification layers. The pre-trained layers were frozen to prevent them from being updated during training. The Adam optimizer was utilized for bettering the model and the training was done with an epoch size of 100.

**D. AMDNet23:**

AMDNet23 Combining CNNs which are convolutional neural systems and long-term short-term memory networks (LSTM) is designated as CNN-LSTM where the strengths of CNNs in image feature extraction are combined with the temporal modeling capabilities of LSTM networks[45]. Before feeding the images to the model, employed a diverse set of augmentation techniques to enhance the training data. These included randomized horizontal and vertical flipping through a probability of fifty percent each and applied random brightness adjustments by varying the brightness level within a range of -0.1 to +0.1. To further increase variation, used random contrast adjustments with factors ranging from 0.8 to 1.2, as well as random saturation adjustments within the bounds of 0.8 to 1.2. Moreover, introduced random hue adjustments to add subtle color variations. Lastly, to augment the dataset further, performed translation-based width and height shifting with a specific range to the input images. The augment strategy emphasizes data diversity and improves the model's broad ability to diagnose unknown data[46]. The model is designed with a depth of 23 layers.

**Input Layer:** The input to the AMDNet23 model is a collection of eye images captured from patients. The input images are represented as a tensor X with dimensions (N, W, H, C), where N corresponds to the eye image's number, and W and H represent The width and length of the images(The model received imagery that measured 256 X 256 in size.) respectively. C denotes the number of color channels in the eye images. This tensor X is then passed into the model's input layer.

**Convolutional Layers (CNN):** After the initial input layer, the CNN component of the model comprises multiple convolutional layers[47]. Every individual layer of convolution utilizes a set of adjustable filters to process the input images. The resulting Features of the map from the $i^{th}$ convolutional layer are denoted as $F_i$, with $i$ ranging from 1 to $n$. The output feature map $F_i$ can be computed as follows:

$$F_i = Conv2D(X, W_i) + b_i$$

Where Conv2D refers to the convolution operation, $W_i$ represents the trainable parameters (weights) specific to the $i^{th}$ convolutional layer, and $b_i$ represents the corresponding biases associated with that layer. The output feature maps $F_i$ have spatial dimensions (W', H') and C' channels.

The model contained six convolutional blocks. The first four convolutional blocks consisted of 2 convolutional layers and 1 batch normalization layer each. The filters were 32, 64, 128, and 256 for the first four blocks, where the kernel size was 3 X 3. The fifth and sixth convolutional blocks consisted of 3 convolution layers and 1 batch normalization layer. All the convolution layers of the fifth and sixth blocks consisted of 512 filters.

**Pooling Layers:** After the layers based on convolution, layers of pooled [48] are frequently used to downsample the feature maps. Let us denote the output feature maps after pooling as $P_i$, where $i$ ranges from 1 to $p$ (the total number of pooling layers). Each pooling layer performs a downsampling operation on the input feature maps. After applying all pooling layers, the resulting feature maps can be denoted as $P_p$ and have spatial dimensions (W", H") and C" channels. The pool size for max-pooling layers was 2 X 2 for all the convolutional blocks.

**LSTM Layer:** The system conveys the development and significance of networks having long-term short-term memory (LSTM), an advancement residing in conventional Recurrent neural networks to delve into the motivation behind LSTM's creation, specifically to address the vanishing gradient problem, which formerly hindered the effective training of RNNs on long sequences [49]. Behind LSTM it introduces memory cells, enabling the network to retain information over extended periods. The mechanism empowers LSTMs in effectively capturing long-term dependencies within the input data. The cell state adds a long-term memory to flows the entire sequence [50]. It enables information to be retained or discarded selectively utilizing the input entrance, forget gatekeeper, and output gateway, which constitute the three main gates. The LSTM cell computations can be mathematically represented as follows, where $t$ denotes the current time step, $x_t$ rrepresents what was the input entered at time $t$, $h_t$ denotes the previous hidden stated, and $c_t$ signifies the cell state:

$$i_t = \sigma(W_i[x_t, h_{t-1}] + b_i)\ldots(1)$$

$$C_t = tanh(W_c[x_t, h_{t-1}] + b_c)\ldots(2)$$

$$C_t = f_t C_{t-1} + i_t C_t \ldots(3)$$

The input gate (1) uses a sigmoid function to combine the previous output $h_{t-1}$ and the present time input $x_t$, deciding the proportion of information to be incorporated into the cell state and (2) employes to obtain new information through the tanh layer to be added into current cell state $C_t$. The current cell state $C_t$, and long term information $C_{t-1}$ are combination into $C_t$(3) whereas $w_i$ determines the sigmoid output and $C_t$ determines to tanh output. "Forget" gate (4) investigates how much of the previous cell state should be retained and carried over to the next time step by assessing probability where $W_f$ and $b_f$ refers to the offset and weight matrix and offset respectively.

$$f_t = \sigma(W_f[x_t, h_{t-1}] + b_f)\ldots(4)$$

The output gate of the LSTM investigates by $h_{t-1}$ and $x_t$ inputs following (4) and (5) passed through the activation function to determine what portion of information to be appeared from the current LSTM unit at timestamp t.

$$o_t = \sigma(W_o[x_t, h_{t-1}] + b_o)\ldots(5)$$

$$h_t = o_t tanh(C_t)\ldots(6)$$

In the above equation, $W_o$ refers to the matrices of the output gate and $b_o$ refers LSTM bias respectively.

**Output Layer:** The output layer delivers the ultimate prediction regarding the existence or non-existence of AMD in the input eye images. We can represent the input tensor to the output layer as $H_{out}$, obtained by reshaping $H_{lstm}$ to have dimensions (NT, D). A function of activation throughout softmax is positioned following its dense layer to the output section. The dense layer takes the input $H_{out}$ and transforms it to generate the output tensor Y, which has dimensions (NT, K). Here, K specifies the number of output classes the model is classifying.

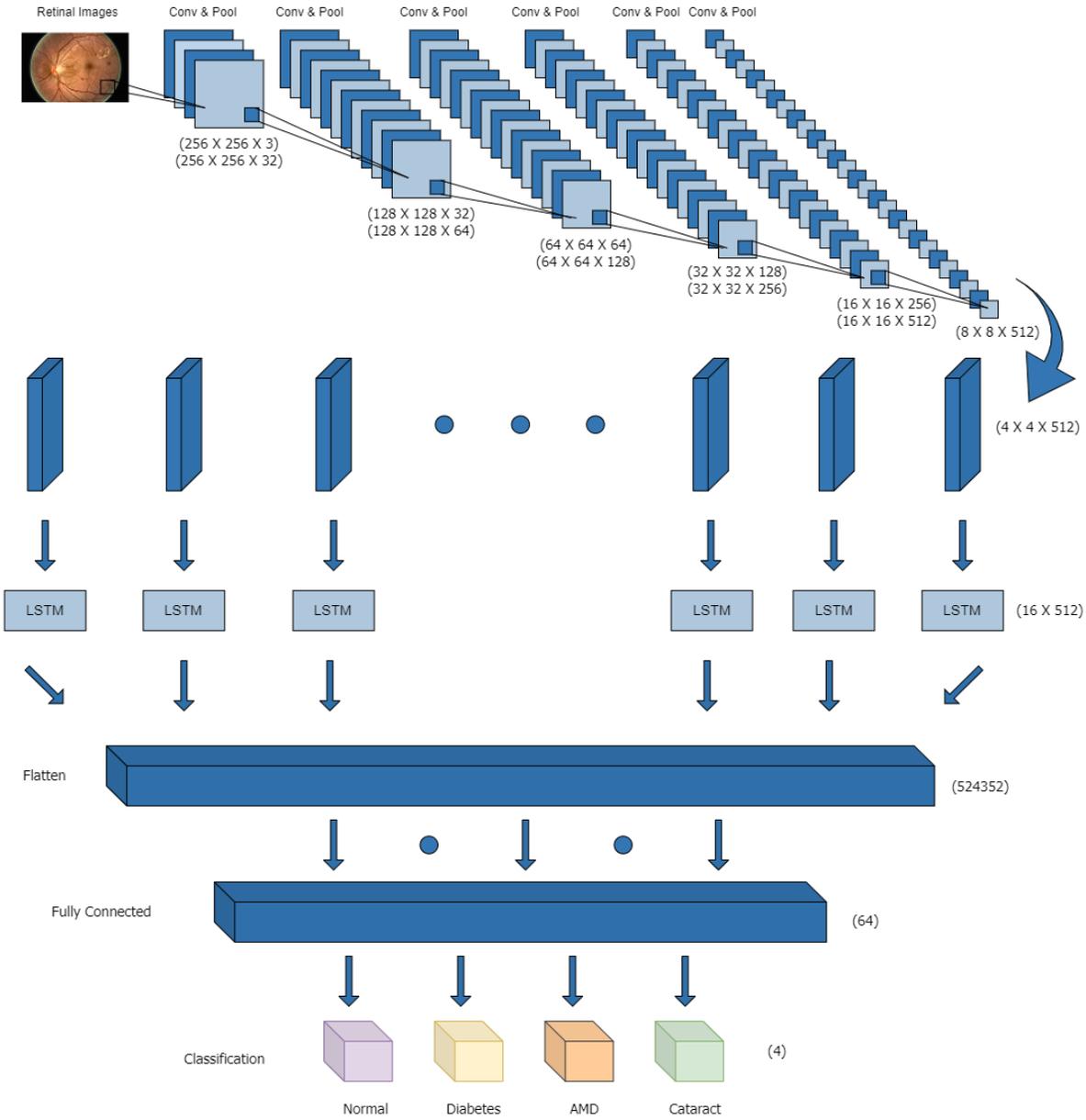

Figure 7: CNN-LSTM system

The AMDNet23 model (Figure 7) for AMD ocular disease detection leverages the complementary strengths of CNNs in spatial feature extraction and LSTMs in modeling sequential dependencies.

| Layer | Type | Kernel Size | Kernel | Input Size |
| --- | --- | --- | --- | --- |
| 1 | Convolution2D | 3 X 3 | 32 | 256 X 256 X 3 |
| 2 | Convolution2D | 3 X 3 | 32 | 256 X 256 X 32 |

| 3 | Maxpooling2D | 2 X 2 | - | 256 X 256 X 32 |
| 4 | Convolution2D | 3 X 3 | 64 | 128 X 128 X 32 |
| 5 | Convolution2D | 3 X 3 | 64 | 128 X 128 X 64 |
| 6 | Maxpooling2D | 2 X 2 | - | 128 X 128 X 64 |
| 7 | Convolution2D | 3 X 3 | 128 | 64 X 64 X 64 |
| 8 | Convolution2D | 3 X 3 | 128 | 64 X 64 X 128 |
| 9 | Maxpooling2D | 2 X 2 | - | 64 X 64 X 128 |
| 10 | Convolution2D | 3 X 3 | 256 | 32 X 32 X 128 |
| 11 | Convolution2D | 3 X 3 | 256 | 32 X 32 X 256 |
| 12 | Maxpooling2D | 2 X 2 | - | 32 X 32 X 256 |
| 13 | Convolution2D | 3 X 3 | 512 | 16 X 16 X 256 |
| 14 | Convolution2D | 3 X 3 | 512 | 16 X 16 X 512 |
| 15 | Convolution2D | 3 X 3 | 512 | 16 X 16 X 512 |
| 16 | Maxpooling2D | 2 X 2 | - | 16 X 16 X 512 |
| 17 | Convolution2D | 3 X 3 | 512 | 8 X 8 X 512 |
| 18 | Convolution2D | 3 X 3 | 512 | 8 X 8 X 512 |
| 19 | Convolution2D | 3 X 3 | 512 | 8 X 8 X 512 |
| 20 | Maxpooling2D | - | - | 8 X 8 X 512 |
| 21 | LSTM | - | - | 16 X 512 |

| 22 | FC | - | 64 | 524352 |
| 23 | Output | - | 4 | 260 |

In this research, An innovative and novel technique was devised to automatically detect AMD by leveraging four distinct types of fundus images. This unique architecture synergizes the power of Neural Networks of Convolutional and Long Short-Term Memory. The CNN module is administered for extracting intricate features from fundus imaging, and the LSTM module serves as the classifier. The proposed AMDNet23 hybrid network for AMD detection consists of 23 layers: It includes 14 convolutional layers placed and 6 layers used for pooling,one fully interconnected layer of (FC), a layer of LSTM and a single output layer with a sense of softmax functionality. In our construction, an individual convolution block is made comprising between two or three 2-dimensional CNN's, A layer with a level of pooling and a layer comprising a twentieth percent dropout rate have of dropouts. Utilizing a convolutional layer with 3x3 kernels and the ReLU function, the feature extraction is carried out efficiently. The input image undergoes dimension reduction using A layer for maximum pooling of $2 \times 2$ kernels. The resulting output structure was discovered (none, 4, 4, 512). the input size inside the layer of LSTM transforms to (16, 512) whenever incorporating the reshaped approach. Combining these two neural network architectures allows the model to effectively analyze eye images, capturing both local spatial patterns and temporal relationships, ultimately enabling accurate AMD diagnosis. The summarized architecture is presented in Table 2.

**Evaluation Criteria**

In this study, as many as 13 models were experimented and the evaluation of all those models will be presented in this section of the paper. Considering the following evaluation criteria, the performance, reliability, and clinical relevance of a AMD detection system can be assessed and also can be determined its suitability for assisting medical professionals in accurately detecting and diagnosing AMD.

**Accuracy**: The accuracy of the AMD detection system in correctly classifying images as AMD, diabetes, cataracts is a crucial evaluation criterion. It evaluates the correctness of the system's detection computed overall.

$$Accuracy = \frac{TP + TN}{TP + TN + FP + FN}$$

**Precision:** The proportion of properly identified AMD situations is examined to measure precision out of all predicted AMD cases.

$$Precision = \frac{TP}{TP + FP}$$

**Sensitivity and Specificity**: Sensitivity is typically referred to by the term the true positive rate, which gauges the system to identify AMD cases correctly. True negative rate, which is often referred to as specificity, assesses its capacity of identifying non-AMD conditions. Both metrics provide insights into the system's performance in different classes and help assess its ability to avoid false positives and false negatives.

$$Sensitivity = \frac{TP}{TP + FN}$$

$$Specificity = \frac{TN}{TN + FP}$$

**F1 Score**: The F1 score offers a comprehensive measurement that addresses the balance between precision and memory and thus represents a harmonious average of precision and recall. It is advantageous in realities whereby there occurs a disparity in class or in cases when the costs of false positives and false negatives fluctuate.

$$F1 Score = \frac{TP}{TP + \frac{1}{2}(FP + FN)}$$

When evaluating model performance, it is crucial to consider a combination of these evaluation criteria accordance with the precise specifications of the completion and the area of expertise. Selecting appropriate metrics and interpreting the results will help determine the effectiveness and suitability of the model. Table 2 showcases the retained value for these evaluation metrics:

| Model | Accuracy | Precision | Sensitivity | Specificity | F1 Score |
|---|---|---|---|---|---|
| ViTB16 | 95.25% | 95.26% | 95.25% | 98.66% | 95.24% |
| MobileViT_XXS | 83.00% | 82.58% | 83.00% | 98.28% | 82.41% |
| InceptionResNetV2 | 47.50% | 36.26% | 47.50% | 82.67% | 40.50% |
| EfficientNetB7 | 92.75% | 92.94% | 92.75% | 98.97% | 92.62% |
| EfficientNetB6 | 92.75% | 92.74% | 92.75% | 97.00% | 92.69% |
| DenseNet121 | 82.25% | 82.41% | 82.25% | 95.00% | 81.46% |
| DenseNet169 | 81.25% | 81.24% | 81.25% | 92.74% | 80.31% |
| DenseNet201 | 84.75% | 84.45% | 84.75% | 93.46% | 84.52% |
| InceptionV3 | 72.25% | 71.53% | 72.25% | 90.16% | 70.71% |
| MobileNetV2 | 71.75% | 71.68% | 71.75% | 88.01% | 71.19% |
| VGG16 | 89.75% | 89.82% | 89.75% | 94.68% | 89.78% |
| VGG19 | 89.00% | 88.96% | 89.00% | 95.62% | 88.80% |
| ResNet50 | 93.25% | 93.37% | 93.25% | 96.73% | 93.17% |
| **AMDNet23** | **96.50%** | **96.51%** | **96.50%** | **99.32%** | **96.49%** |

In conclusion, the AMDNet23 model for AMD ocular disease detection demonstrated strong performance in accurately identifying AMD from eye images. Its high accuracy, precision, and recall values, along with the robust AUC-ROC score, validate its potential as a reliable tool for early detection and intervention. The model's efficiency makes it suitable for practical deployment in healthcare settings, contributing to improved patient care and timely treatment of AMD retinal disease.

**Result & Discussion:**

In this undermentioned portion, the findings of the proposed mechanism along with the comparison with some cutting-edge studies will be comprised. The collected data were divided into sets for conducting training and testing to construct and examine the proposed system. This approach was initially trained to leverage 80% of the data and evaluated utilizing 20% of the collected information. To ensure enhanced productivity several sets of parameters were experimented. The parameter setting that provided us with the most advantageous outline for the proposed model is given below

| | |
|---|---|
| **Batch size** | 32 |
| **Epochs** | 100 |
| **Learning rate** | 0.001 |
| **Decay rate** | 0.95 |
| **Decay step** | 1 |

Figure 5 and Figure 6 show both training and test sets of data, the accuracy and loss curves. The graphical representations of epoch versus accuracy and epoch versus losses are valuable insights for monitoring and understanding the progress of the proposed method.

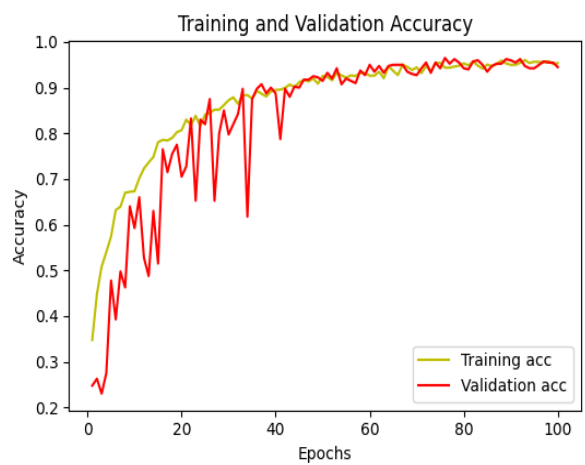 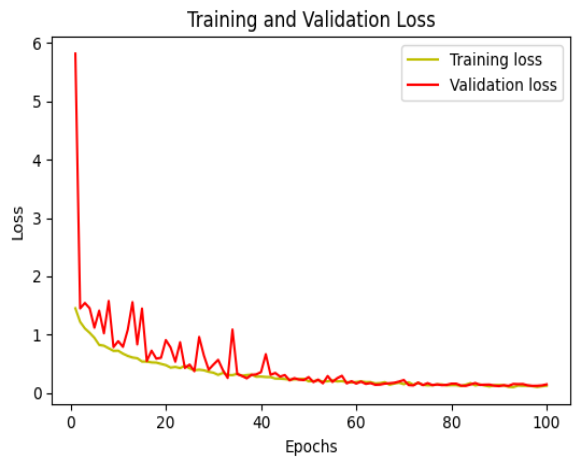

The graph exhibits how the model's accuracy varies while an increasing amount of training epochs raises. The predictability of the model on either a set of training data or a testing set appears on the axis in the vertical direction, along with the number of epochs denoted throughout the horizontal direction. It is observed from Figure 5 reveals the overall number of epochs rises over training, the model's accuracy elevates as it learns from the training data. At first, the accuracy continues to improve with each epoch, it suggests that the model can benefit from additional training. Later the accuracy eventually plateaus. This indicates that the model has converged and further training may not significantly improve accuracy.

The epoch vs loss graph demonstrates the association between the process of training number of epochs and the loss or error of the model. The loss of performance is a disparity among the estimated of model outline and the intended outline. The loss level is portrayed through the y-axis, whereas the total amount of epochs is displayed through the x-axis. In Figure 6, the loss is seen initially high as the model makes random predictions. The training messages, the loss decreases, reflecting the model's improved performance and ability to make more accurate predictions.

Figure X a represents The AMDNet23 model's confusion matrix, resulting in was determined on the basis of the ablation investigation, Adam, and learning rate, experiences the greatest level of accuracy and is configured in a great potential manner.

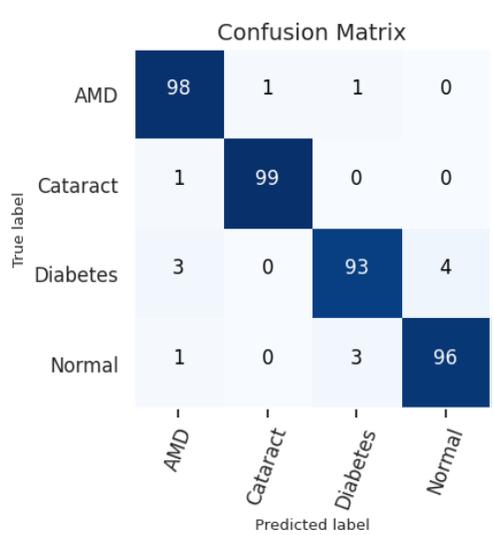

The row of values reflects what is actually labeled attached to the images, while the column-specific values reveal the quantities provided the predictive model estimates. The diagonal values indicate correct predictions (TP). However, the model had the great results for AMD. According to the confusion matrix, Two of the AMD pictorials was mistakenly classified as cataract-related and diabetes, whereas 98 among the 100 AMD investigations possessed effectively diagnosed. Next, 99 among a possible 100 anticipated involving cataracts taught correctly where one as AMD. Of 100 diabetes images, Seven images had been mistakenly identified, involving 3 referred to as AMD along with 4 as belonging to the healthy class, exposing of 93 correctly assigned. In closing least, among the 100 normal images, 96 were perfectly identified, and 1 had been misinterpreted as images related to AMD and 3 associated to diabetes.

**State-of-the-art work comparison**

Table 3 provides a concisely summarizes the main approaches in the existing literature for diagnosing AMD disease and our proposed method. These approaches primarily involve conventional methods and deep learning algorithms, which utilize retinal images for diagnosis.

| Author | Year | Method | No. of images | Accuracy |
|---|---|---|---|---|
| TK Yoo et. al. [51] | 2018 | VGG19-RF | 3000 | 3- Class 95% accuracy |
| Huiying Liu et. al [52] | 2019 | DeepAMD | 4725 | 6- Class 70% accuracy |
| Felix Grassman et. al. [53] | 2020 | Ensemble networks net | 3654 | 13- class 63% accuracy |
| N Chea and Y Nam [54] | 2021 | Optimal residual deep neural networks | 2335 | 4- Class 85.79% accuracy |
| C Domínguez et. al [55] | 2023 | Transformer-based system | 4896 | 3-Class 82.55% accuracy |
| P Zang et. Al [56] | 2023 | Deep-Learning based aided system | Not specified | 4-Class 80% accuracy |
| **Proposed Method** | **2023** | **AMDNet23** | **2000** | **4- Class 96.5% accuracy** |

The AMDNet23 model proposed in the study achieves a high accuracy rate of 96.5%, surpassing other state-of-the-art works currently available. As a result, It can be presumed that this proposed method is effective for early-stage detection and diagnosis of AMD, and this novel method also diagnoses Cataracts and diabetic retinopathy utilizing fundus ophthalmology datasets, demonstrating superior accuracy.

**Comparison of the AMDNet23 model with the transfer Learning models:**

The outline demonstrated and effectively potential of hybrid AMDNet23 network for precisely detecting AMD eye disease from images. The capacity of the model to precisely detect AMD instances is demonstrated by the excellent precision, accuracy, recall and F1-score acquired. Combination of CNNs and LSTMs allows for the extracting of both spatial and temporal features, capturing the subtle patterns and changes associated with AMD. Figure 5 exhibits the comparison of performance between the model we proposed and a several pre-trained prepared.

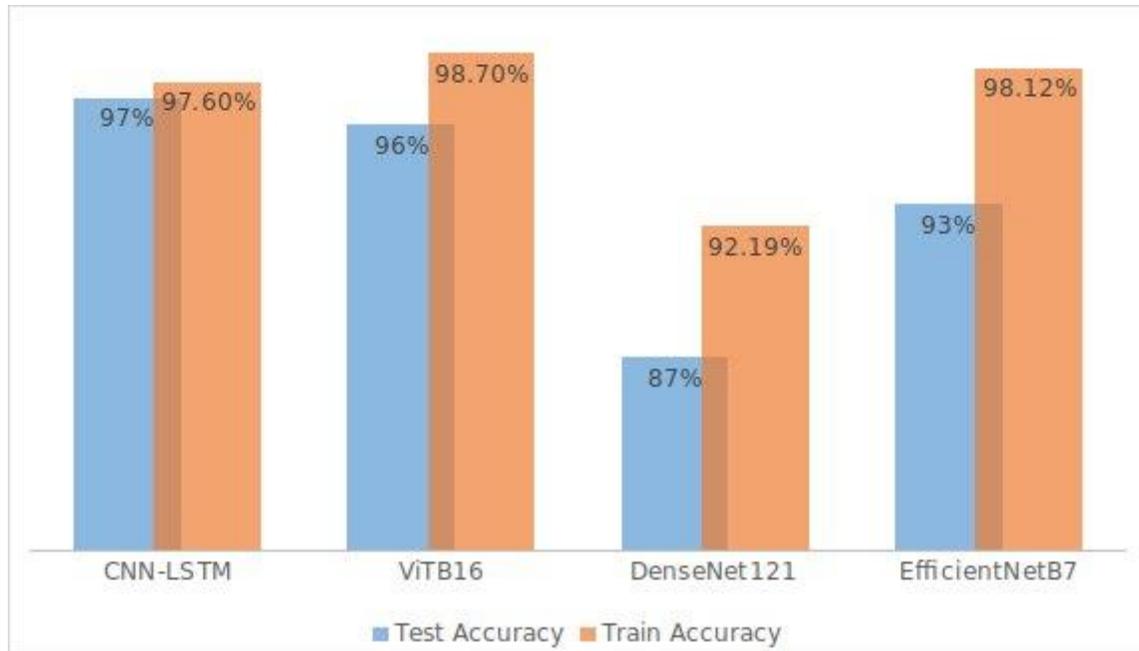

**Conclusion**

In essence, this study proposed a AMDNet23 model for detecting and diagnosing AMD disease using several image datasets. The model achieved a high accuracy rate of 96.5%, surpassing other state-of-the-art works in the field. Furthermore, when compared with pre-trained models, the novel deep AMDNet23 method also showed superior accuracy for AMD detection, and the system is efficient to diagnose cataracts and diabetic retinopathy respectively. In the future, incorporating additional modalities or features can potentially enhance the performance of AMD detection models. Combining fundus images with other clinical data, which could include patient demographics, health records, or genetic information, may improve accuracy and enable a more comprehensive awareness of the disease. In broadly, the findings of this research clearly demonstrates effectively of the proposed AMDNet23 model in accurately detecting and diagnosing AMD cases. This model holds promise for early detection and diagnosis of AMD ocular disease, which could assist clinicians and aid in timely intervention and treatment for affected individuals.